\documentclass[12pt]{article}%
\usepackage{amsfonts}
\usepackage{mitpress}
\usepackage{amsmath}
\usepackage{amssymb}
\usepackage{graphicx}%
\providecommand{\U}[1]{\protect\rule{.1in}{.1in}}
\DeclareGraphicsExtensions{.pdf,.eps,.ps,.png,.jpg,.jpeg}
\providecommand{\U}[1]{\protect\rule{.1in}{.1in}}

\newdimen\dummy
\dummy=\oddsidemargin
\begin{document}

\title{Bosons, Fermions, Spin, Gravity, and the Spin-Statistics
Connection\thanks{Invited article written for the S. N. Bose 125-birth
anniversary commemorative issue of Physics News (Indian Physics Association).}}
\author{C. S. Unnikrishnan\\Tata Institute of Fundamental Research, Mumbai 400005, India}
\maketitle

\begin{abstract}
Satyendra Nath Bose's attempt to describe the quantum statistical aspects of
light consistently in terms of particles, and Einstein's generalisation, lead
to the concept of Bosons as a class of quanta obeying `Bose-Einstein
statistics'. Their identity as a class came in sharp contrast when the Pauli
exclusion principle and the Dirac equation revealed the other class called
Fermions, obeying  `Fermi-Dirac statistics'. Spin, and spin alone, is the
determining factor of the multiparticle behaviour of fundamental quanta. This
is the basis of the Spin-Statistics Connection. While it is known that the
overall theoretical picture is consistent, the physical reason for the
connection is unknown. Further, the class difference is sensitive only to the
total spin in a quantum aggregate, as spectacularly seen in superconductivity
and superfluidity, and in the Bose-Einstein condensation of neutral atomic
gas. Can we grasp the true reason behind the difference in the collective
behaviour of Bosons and Fermions? An explorer's journey demanding logical and
physical consistency of what we already know takes us to the hidden factors in
the relation between spin and the statistics of quanta. The surprising answer
is in the domain of gravity, that too, on a cosmic scale.

\end{abstract}

\section*{Prologue}

The fundamental particles of nature are just two kinds as far as their
collective quantum dynamical behaviour is concerned, governing much of the
physical phenomena of the macroscopic world. While any number of identical
particles with integer spin, called Bosons, can occupy a physical state, only
one particle per state is possible for identical particles with half-integer
spin, called Fermions. The contrasting statistical property of both kinds of
particles were first phenomenologically introduced, notably by S. N. Bose for
the integer spin photons, and by W. Pauli, for the half-integer spin electrons
in atoms. A connection between the spin and `statistics' was established by
Pauli in terms of the consistency of the underlying relativistic quantum field
theories \cite{Pauli1940} but there is no physical understanding of these
rules to date. The lack of true understanding is seriously felt when one
recognizes that these rules, stated in the context of relativistic field
theories, are operative in nonrelativistic dynamics, even for just a pair of
particles of a kind. The frustration about the unsatisfactory situation was
expressed in a comprehensive book on the spin-statistics connection by Ian
Duck and George Sudarshan \cite{Sud-Duck}:

\begin{quote}
Everyone knows the spin-statistics theorem, but no one understands it... What
is proved - whether truly or not, whether optimally or not, in an acceptable
logical sequence or not - is that the existing theory is consistent with
spin-statistics relation. What is not demonstrated is a reason for the
spin-statistics relation.
\end{quote}

This lament is not new. Feynman admitted in his `Lectures in physics'
\cite{Feynman},

\begin{quote}
Why is it that particles with half-integral spin are Fermi particles whose
amplitudes add with the minus sign, whereas particles with integral spin are
Bose particles whose amplitudes add with the positive sign? We apologize for
the fact that we cannot give you an elementary explanation. An explanation has
been worked out by Pauli from complicated arguments of quantum field theory
and relativity. He has shown that the two must necessarily go together, but we
have not been able to find a way of reproducing his arguments on an elementary
level... . This probably means that we do not have a complete understanding of
the fundamental principle involved.
\end{quote}

Though the connection is called Spin-Statistics Connection (SSC), it is not a
`statistical' connection because the rules are effective for just two
identical nonrelativistic particles. The bipartite behaviour then influences
the entire ensemble and their statistical behaviour. Another point to note is
that the formal wavefunctions for a pair of Bosons (and Fermions) need to
specify their symmetry (anti-symmetry) property even when the particles are at
space-like separations, anticipating the possibility that two particles can
approach each other in their future. Finally, the human need to make
consistent theories and understand the physical world cannot be the
\emph{reason} for a fundamental fact in the physical world! If we follow
Bose's attitude about the need to have uncompromising logical integrity in
dealing with physical problems, we cannot be satisfied with the present
situation of the lack of understanding of the spin-statistics connection. The
entire confidence in the formal proofs of SSC rests on the absolute
correctness of the special theory of relativity. In fact, Pauli's 1940 paper
\cite{Pauli1940} ends with the statement,

\begin{quote}
In conclusion we wish to state that according to our opinion the connection
between spin and statistics is one of the most important applications of the
special relativity theory.
\end{quote}

The special theory requires that space is empty and remains isotropic in every
moving frame. In anisotropic spaces one cannot preserve a isotropic Minkowski
metric. However, a fact realized three decades after the special theory shows
that the factual situation is very different; we live in matter-filled
universe, instead of largely empty space. This renders space and its metric
anisotropic in moving frames, as evidenced clearly in observational cosmology.
We choose to ignore this fatal inconsistency, despite clear experimental
proofs. This inconsistency affects the entire edifice \cite{Unni-Shyam} and we
need to reconstruct a logically consistent and physically reasoned proof for
the SSC. This goal cannot be achieved without a proper understanding of the
true physical relevance of `spin' itself.

\emph{Spin is the closed localized current of the charge of gravity, or
mass-energy}. Thus, spin is to gravity as the magnetic moment is to
electromagnetism. Just as every physical effect involving a magnetic moment is
traceable to its interaction with an electromagnetic field, and hence its
sources, every spin related physical effect should have a connection to a
gravitational field. These are the only two long range fields in this world
and all physical explanations have to rely on these interactions. The
difference between a pair of Bosons and a pair of Fermions is the phase of the
wavefunction under exchange of their spatial positions. As we will see,
exchange in the presence of sufficient amount of matter generates exactly such
a phase, due to the gravitational interaction. Where is the gravitational
field large enough to be capable of affecting a microscopic spin? It turns out
that the gravity of the entire matter-energy of the observed universe is just
right and large to provide spin-dependent quantum phase that is the crucial
physical input to understand the spin-statistics connection. Thus, \emph{the
physical reason for SSC is this cosmic gravitational connection}. Before we
discuss this profound link, we will review several important aspects related
to the spin-statistics connection.

\section{Bose, Einstein and Bosons}

The important contribution of S. N. Bose to the understanding of the Planck
radiation formula, in Bose's own words, was \textquotedblleft the derivation
of the factor $8\pi\nu^{2}/c^{3}$ without using any classical aspects of
radiatio\textquotedblright. The derivation \cite{Bose1924} motivated by
consistent pedagogy and logical integrity, went much further and contained
implicitly the concept of indistinguishability of identical quanta. The
important aspects of the indistinguishability of identical particles had been
already used as a crucial input in the statistical mechanics of atoms, by J.
W. Gibbs in his resolution of the Gibbs paradox \cite{Jaynes,Unni-Gibbs}. The
often repeated statements that Bose introduced the concept of
indistinguishable identical particles is incorrect, as clear from Bose's
second paper in 1924, where he wrote \cite{Bose1924-2},

\begin{quote}
Debye has shown that Planck's law can be derived using statistical mechanics.
His derivation is, however, not completely independent of classical
electrodynamics, because he uses the concept of normal modes of the ether and
assumes that for calculating the energy the spectral range between $\nu$ and
$\nu+d\nu$ can be replaced by $8\pi V\nu^{2}/c^{3}$ resonators whose energy
can be only multiples of $h\nu$. One can however show that the derivation can
be so modified that one does not have to borrow anything from the classical
theory. $8\pi V\nu^{2}/c^{3}$ can be interpreted as the number of elementary
cells in the six dimensional phase space of the quanta. The further
calculations remain essentially unchanged.
\end{quote}

(The words `identical' and `indistinguishable' do not appear in Bose's paper).

Einstein's involvement in Bose's work was a chance event, due to Bose's
confident correspondence with him on the importance and uniqueness of his
derivation that eliminated the logical deficiency in the derivations by Debye
and Einstein. But Einstein's immediate interest in the derivation was due to
his noticing the possibility of its generalization to the quantum theory of a
monatomic gas. He did this with due praise to Bose's method, with the
spectacular prediction of the non-intuitive `condensation', now known as the
Bose-Einstein condensation \cite{Einstein-BEC}. This was a condensation
`without interaction' into a zero-pressure state with no kinetic energy. The
role of the spin was not realized because the quantum spin was not yet part of physics.

Bose treated the \emph{quanta of light as real particles} with energy $E=h\nu$
and momentum $p=h\nu/c$. He thought that using the waves of radiation, a
concept belonging to classical electrodynamics, in the quantum theory of
particles was inconsistent and logically flawed. L. de Broglie's daring
proposal of wave-particle duality came in late 1924, and the idea was known
only to a few, like Einstein. Bose argued that the volume of the phase space
with momentum between $p$ and $p+dp$, occupying spatial volume $V$, would be
$4\pi Vp^{2}dp=4\pi h^{3}\nu^{2}d\nu/c^{3}$. He assumed that the light quanta
occupied `elementary cells' of volume $h^{3}$ in the phase space of
coordinates and momenta. Accounting for the two `polarization' states and
dividing by volume of each phase space cell ($h^{3}$), one gets the correct
weight factor $8\pi Vh^{3}\nu^{2}/c^{3}$ in the radiation spectrum. Thus, Bose
derived the factor `quantum theoretically', \ \emph{for the first time}. By
avoiding the modes of waves, Bose's derivation marked an important conceptual
advance. Bose had even considered the possibility of a two-valued helicity for
photons in his derivation of the Planck spectrum. Apparently, Bose meant and
wrote `two states of spin directions' rather than `polarization states', but
Einstein changed it to a more conventional, though inconsistent, description
\cite{PGhose-Bose,Blanpied}. In any case, the importance of spin in the
collective behaviour of particles became clear by 1926, with the clear
division into the two classes of Bose-Einstein statistics for integer-spin
particles and Fermi-Dirac statistics for half-integer spin particles. It was
Dirac who named the integer class as Bosons. The associated quantum field
theories used tensor fields for the former and spinor field for the latter.
Thus, spin, statistics and the transformation properties under spatial
rotations were all linked.

\section{Fermi, Fermions and their Statistics}

Pauli's exclusion principle, proposed in 1925, was a prohibitive principle
distilled from atomic data, advanced as a decree that bans electrons with the
same quantum numbers to occupy the same orbital shell \cite{Pauli-exclusion}.
The `fourth quantum number' of pre-mature quantum mechanics was introduced as
a two-valued quantity, which eventually was identified as the two projections
of the half-integer spin. The proposal by Uhlenbeck and Goudsmit
\cite{Uhlen-Goud} connecting the fourth quantum number and the intrinsic spin
of the electron was not easy to digest, though plausible. A classically
indescribable two-valuedness of a familiar physical quantity like angular
momentum that is orientable freely in 3-dimensional space was a difficulty to
incorporate in the new quantum theory. That it was genuine angular momentum,
and its tight relationship between the spin and the magnetic moment, was
indisputable in experimental spectral data. The theoretical issues were
cleared and solved soon, mainly by L. Thomas, P. Jordan, W. Pauli and P. Dirac
\cite{Tomonaga}. The two-valuedness of the spin projections even for the
vector field of electromagnetism was understood as linked to the massless
nature of the quanta. In 1926, a year after the genesis of Bose-Einstein
statistics, Enrico Fermi published the equivalent of Einstein's 1924 paper on
the quantum theory of monatomic gases (in the same journal where Bose's 1924
paper appeared), with the Pauli exclusion as a constraint \cite{Fermi1926}. He wrote,

\begin{quote}
It shall in the present work be assumed only the rule, that was proposed for
the first time by Pauli and proved by many spectroscopic facts, that two
equivalent elements can never exist in a system, whose quantum numbers
completely agree. The equation of state and internal energy of the ideal gas
shall be deduced from this assumption.
\end{quote}

It was left for Fermi himself to suggest, a year later, an application for the
electrons in a metal, later known as the Thomas-Fermi model. The
characteristic difference between Bosons and Fermions is seen in the
probability function for their statistical distribution, as a crucial quantum
signature differing in sign, $1/\left(  \exp(\frac{E}{kT})-1\right)  $ for
Bosons and $1/\left(  \exp(\frac{E}{kT})+1\right)  $ Fermions.

\section{Phenomena, from atoms to stars}

The rich variety of elements and nature itself can be traced to the peculiar
behavior of electrons, and other Fermions -- neutrons and protons -- in atoms.
The quantum mechanical nature of Bosons was more subtle and difficult to see,
because any number of identical particles can be in the same state even in
classical statistical mechanics of Boltzmann and Maxwell. Therefore, the
signature quantum feature of Bosons is not that any number of particles can
occupy the same state, but the probability is enhanced if there are already
some particles occupying the state; Bosons tend to bunch together. The special
feature of the \textquotedblleft$-1$\textquotedblright\ factor in the
distribution is evident only at very low temperature, as in the prediction of
`condensation without any interaction' by Einstein. In contrast, the exclusion
principle and Fermionic behaviour were evident in diverse phenomena, from the
electronic transport in metals to behaviour of `finished' stars. R. H. Fowler,
who presented Dirac's paper on the new statistics in the Royal Society,
applied the wisdom to dense dwarf stars and resolved the puzzle about their
stability \cite{Fowler-stars}. It is striking that Fowler's description --
\textquotedblleft Its essential feature is a principle of exclusion which
prevents two mass-points ever occupying exactly the same cell of extension
$h^{3}$ in the six dimensional phase-space of the
mass-points\textquotedblright\ -- uses the key idea of `minimal cell'
introduced by Bose for the statistics of photons. The first application of
Fermi statistics to relativistic electrons was S. Chandrasekhar's amazing
discovery of the critical mass of white dwarf stars, beyond which gravity wins
over exclusion \cite{Chandra-limit}. In the next decade, the discovery of
superfluidity in liquid Helium was soon identified by F. London as the
phenomenon of Bose-Einstein condensation \cite{London-BEC}. This confirmed
that Fermionic composites, like the Helium atom consisting of two Fermions
each of protons, neutrons and electrons, behave as a Boson. What matters for
statistics is not the fundamental fields, but the gross spin of the assembled
entity. By then, quantum theory of particles and fields, and their numerous
applications in various physical problems of the micro-world were well
charted. Superconductivity was identified as the Bose-Einstein condensation of
paired electrons, called Cooper pairs. The BCS theory was further confirmation
that even loosely bound pair of Fermions behave as Bosons, due to their net
integer spin \cite{BCS}. The next half of the century saw spectacular
discoveries like masers and lasers (Bosons) and neutron stars (Fermions). The
quantum discoveries and their theories involving multi-particle quantum
effects seem never ending.

\section{Pauli's struggle and triumph}

The two-valued quantum number and exclusion principle was introduced into
atomic physics in 1924-25 by Pauli. This was in the context of the spectrum of
Hydrogen-like alkali atoms. He became obsessed with the need to establish the
reasons for his exclusion principle, and it took over much of his intellectual
and psychological space for two decades \cite{Pauli-137}. Rather than trying
to find the physical reason for the exclusion tendency of Fermions, the quest
went in another direction, of finding the reasons for the difference between
Fermions and Bosons. This seemed feasible since the difference in the
collective behaviour was linked to whether the spin was integer valued or
half-integer valued. All known quantum fields belonged to these two classes.
The integer spin fields had scalar, vector, tensor representations and the
half integer fields were spinors. The characteristic theoretical difference
was encoded in the algebra of the creation and annihilation operators that
linked fields to the particles in the theory. Pauli said in his Nobel lecture
of 1945 \cite{Pauli-Nobel},

\begin{quote}
In order to prepare for the discussion of more fundamental questions, we want
to stress here a law of Nature which is generally valid, namely, the
connection between spin and symmetry class. A half-integer value of the spin
quantum number is always connected with antisymmetrical states (exclusion
principle), an integer spin with symmetrical states. This law holds not only
for protons and neutrons but also for electrons. Moreover, it can easily be
seen that it holds for compound systems, if it holds for all of its
constituents. If we search for a theoretical explanation of this law, we must
pass to the discussion of relativistic wave mechanics, since we saw that it
can certainly not be explained by non-relativistic wave mechanics.
\end{quote}

If $\psi_{1}(x)$ and $\psi_{2}(x)$ are the wavefunctions for the two
particles, the two-particle symmetric wavefunction is $\Psi_{s}(x_{1}%
,x_{2})=\psi_{1}(x_{1})\psi_{2}(x_{2})+\psi_{2}(x_{1})\psi_{1}(x_{2})$ and the
anti-symmetric wavefunction is $\Psi_{a}(x_{1},x_{2})=\psi_{1}(x_{1})\psi
_{2}(x_{2})-\psi_{2}(x_{1})\psi_{1}(x_{2})$. This difference is sign captures
the Pauli exclusion, since the wavefunction vanishes if the states are
identical with the same coordinates. This was made clear by Jordan and Wigner
in an article in 1928 \cite{Jor-Wig}. They compared Bose-Einstein situation
with Pauli's (Fermi-Dirac) and stated, \textquotedblleft We can say that the
existence of material particles and the validity of the Pauli principle can be
understood as a consequence of the quantum mechanical multiplication
properties of the de Broglie wave amplitudes.\textquotedblright

The field amplitudes in the classical theory, $a$ and its complex conjugate
$a^{\ast}$, become the annihilation operator $\hat{a}$ and the creation
operator $\hat{a}^{\dag}$ in the quantum theory. The operator $N=a^{\dag}a$ is
the number operator that counts the total quanta (we drop the `operator hats).
Hence, the energy at frequency $\nu$ is $E=h\nu(a^{\dag}a+1/2)$. The algebra
obeyed by these operators is known to be that of the `commutator',
\begin{align}
\lbrack a,a^{\dag}]_{-}  &  =aa^{\dag}-a^{\dag}a=1\nonumber\\
\lbrack a,a]_{-}  &  =[a^{\dag},a^{\dag}]_{-}=0
\end{align}
For Fermions, the situation is different. Dirac equation is the basis for
describing massive Fermions. Now there are two sets of creation and
annihilation operators, meant for particles and their anti-particles. Pauli
exclusion principle is implemented by choosing anticommutation algebra for
these operators, as first done by Jordan.%

\begin{align}
\lbrack a,a^{\dag}]_{+}  &  =aa^{\dag}+a^{\dag}a=1\nonumber\\
\lbrack b,b^{\dag}]_{+}  &  =bb^{\dag}+b^{\dag}b=1
\end{align}
All other anticommutators are zero. Then $a^{\dag}a^{\dag}=aa=0$. The only
eigenvalues of the number operators $a^{\dag}a$ and $b^{\dag}b$ are $0$ and
$1$, realizing Pauli exclusion. This has no classical counterparts in the
field amplitudes. However, the number operator is still $N=a^{\dag}a$ and
$\bar{N}=b^{\dag}b$.

I will first quote Pauli's own description \cite{Pauli-progress1950} of what
he achieved in his proof of the SSC:

\begin{quotation}
In relativistically invariant quantized field theories the following
conditions are fulfilled in the normal cases of half-integer spin connected
with exclusion principle (Fermions) and of integer spin connected with
symmetrical statistics (Bosons).

1. The vacuum is the state of lowest energy. So long as no interaction between
particles is considered the energy difference between this state of lowest
energy and the state where a finite number of particles is present is finite.

2. Physical quantities (observables) commute with each other in two space-time
points with a space-like distance. (Indeed due to the impossibility of signal
velocities greater than that of light, measurements at two such points cannot
disturb each other.)

3. The metric in the Hilbert-space of the quantum mechanical states is
positive definite. This guarantees the positive sign of the values of physical
probabilities. There seems to be agreement now about the necessity of all
three postulates in physical theories. In earlier investigations I have shown
that in the abnormal cases of half-integer spin connected with symmetrical
statistics and of integer spin connected with exclusion principle, which do
not occur in nature, not all of the three mentioned postulates can be
fulfilled in a relativistically invariant quantized field theory. In this
older formulation of the theory for the abnormal cases the postulate (1) was
violated for half-integer spins and the postulate (2) for integer spins, while
postulate (3) was always fulfilled.\ (Progress of Theoretical Physics, 1950).
\end{quotation}

Following Sudarshan and Duck \cite{Sud-Duck-AJP}, I indicate the case for
spin-1/2 Fermions. The Dirac wavefunction, for $p\approx0$ is of the form%
\begin{equation}
\chi=a\psi_{+}e^{-imt/\hbar}+b^{\ast}\psi_{-}e^{+imt/\hbar}%
\end{equation}
with $m$ positive. We have $i\hbar a\frac{\partial}{\partial t}e^{-imt/\hbar
}=ame^{-imt/\hbar}$ and $i\hbar b^{\ast}\frac{\partial}{\partial
t}e^{imt/\hbar}=-b^{\ast}me^{imt/\hbar}=Ee^{imt/\hbar}$. In the filed
quantization, amplitudes $a$ and $b^{\ast}$ become the annihilation and
creation operators $a$ and $b^{\dag}$. The energy in the state $\chi$ is
\begin{align}
E  &  =i%
{\textstyle\int}
\chi^{\dag}\frac{\partial}{\partial t}\chi d^{3}x=m(a^{\dag}a-bb^{\dag
})\nonumber\\
&  =m(a^{\dag}a+b^{\dag}b-1)=m\left(  N+\bar{N}-1\right)
\end{align}
If we assume the commutation relation $[a,a^{\dag}]_{-}=aa^{\dag}-a^{\dag}a=1$
and $[b,b^{\dag}]_{-}=1,$ instead of anticommutation, $E=m(a^{\dag}a-b^{\dag
}b-1)=m\left(  N-\bar{N}-1\right)  $, This energy is not positive definite.
Thus, one concludes that Dirac spin-1/2 field cannot be consistently quantized
with commutation relations.

The path taken by Pauli, and established with his authority, has been followed
by many mathematical physicists. As a result, the proofs were made more
formal, mathematically rigorous, and applicable to interacting particles etc.
However, the universality and simplicity of the actual physical situations
where the SSC is applicable have not been captured. \ We reiterate the
features that prompted Sudarshan and Duck, and Feynman to make their comments:
The proof does not reveal the reason for the SSC; there is no physical
insight. The elementary situation of the quantum mechanics of a pair of
identical non-relativistic particles remains as an enigma. Particles are said
to obey their statistics for the consistency of the theories of the
hypothetical fields that represent them. The reasons the `wrong statistics'
violate the requirement of consistency are very different for Bosons and
Fermions. In short, the proof is really unsatisfactory. It was unsatisfactory
even for Pauli! He ended his Nobel lecture thus,

\begin{quote}
At the end of this lecture I may express my critical opinion, that a correct
theory should neither lead to infinite zero-point energies nor to infinite
zero charges, that it should not use mathematical tricks to subtract
infinities or singularities, nor should it invent a `hypothetical world' which
is only a mathematical fiction before it is able to formulate the correct
interpretation of the actual world of physics.

From the point of view of logic, my report on `Exclusion principle and quantum
mechanics' has no conclusion.
\end{quote}

\section{Physical proofs of the spin-statistics connection?}

A study of the work on the the reasons for the spin-statistics connection
leaves anybody, who desires a coherent and physically reasonable understanding
into the most remarkable characteristic of two-particle behaviour,
disappointed. As emphasized, what is shown is that the quantization of the
corresponding fields with the inappropriate algebra of the
creation-annihilation operators leads to inconsistency. Thus, Bosons need
commutation relation; otherwise the requirement that the field operators
commute at space-like separated points cannot be maintained in the theory.
Fermions need anti-commutation relation; otherwise the requirement of the
positivity of the energy cannot be guaranteed, in the theory. The standard
proofs rely on very different reasons for the spin-statistics connection of
Bosons and Fermions.

There is a wide chasm between the physical situations in which the SSC is
operative and the premises of the formal proofs. The SSC is most spectacularly
and commonly seen in many nonrelativistic situations. Does one has to rely on
a proof based on relativistic quantum field theories for proving SSC for the
modest spin-1 phonons in the solid lattice? Even in situations where
fundamental particles are involved, when the Fermions can form pair-wise
bonds, as in the Cooper pairs of superconductivity, the Fermion composites
behave as Bosons. For such composites, one will have to argue that the
effective theory applicable is a Klein-Gordon like field theory, while the
rest of the electrons obey a spinor field theory; that seems far-fetched in
the proof of a fundamental fact like SSC. Finally, I consider that it is very
important to realize that every physical change, including changes in quantum
phase, must occur through a physical interaction involving one of the
fundamental interactions. This point will become obvious as our discussion progresses.

The behaviour of Fermions under pressure is well known in the case of white
dwarfs and neutron stars. The degeneracy pressure can be defeated by gravity
even though the particles remain as Fermions. Pauli would have been very
disappointed and plunged further into depression at the breach of his
exclusion Principle and any negation of the formal proof by overwhelming gravity.

There have been some sustained efforts by many to find a physically reasoned,
and possible `simple', proof of the spin-statistics connection. These tried to
derive the crucial phase factor under exchange of a pair of particles, without
using relativistic physics or mathematical aspects of field theory
\cite{Sud-Duck,Sud-Duck-AJP}. So, they were focusing on the statement by
Jordan and Wigner that the \textquotedblleft validity of the Pauli principle
can be understood as a consequence of the quantum mechanical multiplication
properties of the de Broglie wave amplitudes\textquotedblright. However, these
proofs need additional assumptions and some of these could be circular, with a
hidden connection to the proposition to be proved.

The spin-statistics connection usually stated as

a) Particles with integer spin are Bosons and they obey the Bose-Einstein statistics,

b) Particles with half-integer spin are Fermions and they obey the Fermi-Dirac statistics,

can be restated in the Jordan-Wigner spirit as

a) The quantum amplitude for a scattering event between identical integer spin
particles and the amplitude with an exchange of the particles add with a plus
($+$) sign. In other words, the phase difference between the direct amplitude
and the exchanged amplitude is an integer multiple of $2\pi.$

b) The amplitude for a scattering event between identical half-integer spin
particles and the amplitude with an exchange of the particles add with a minus
($-$) sign. In other words, the phase difference between the direct amplitude
and the exchanged amplitude is an odd integer multiple of $\pi.$

A geometric understanding of these statements was published by Berry and
Robbins \cite{Berry}, and several authors have invoked the relation between
rotation operators and exchange of particles in quantum mechanics while
attempting to prove the spin-statistic theorem \cite{Sud-Duck-AJP}. E. C. G.
Sudarshan has been arguing for the existence of a simple proof that is free of
arguments specific to relativistic quantum field theory \cite{Sud-Duck}. His
proof of the SSC used rotational invariance, in conjunction with the postulate
of flavor symmetry of the Lagrangian. Flavor symmetry is assumed to prevent a
free antisymmetrization on internal `pseudo-spin' degrees of freedom ~like
isospin, which could reverse the conclusions.

The general assessment seems to be that none of the proofs projected as a
simple proof, can be considered as valid proofs. This has been implied in a
critical review of the book by Sudarshan and Duck \cite{Sud-Duck}, by A. S.
Wightman \cite{Wightman}. The grave nature of the situation is clear from the
fact that even the very small fraction of physicists who are experts on these
formal proofs do not agree with each other on their own attempts at more
accessible proofs. Thus, there is no consensus on whether any of these proofs
is taking us closer to a better understanding of SSC, let alone a physical understanding.

\section{Comments on Anyons}

Physics in two spatial dimensions have become very important, with many
situations in condensed matter physics. Topological considerations become
relevant for SSC when dynamics is confined to two dimensions. The quantization
of fundamental particles into two groups of integer and half integer spin
remains intact. However, the statistics for certain composite objects has more
general options. I discuss just one example involving a charged particle and a
`flux tube'. A wavefunction of a charged particle of charge $e$ going a full
circle around a solenoid containing magnetic flux $\phi$ picks up the quantum
phase $e\phi/\hbar$. Therefore, two such particle-flux composites has a net
phase $e\phi/\hbar$ under exchange of their spatial coordinates. Since the
exchange phase is now arbitrary, with any value between 0 and $2\pi$, the
statistics is not restricted to that of either Bosons or Fermions. Such
composites are named `anyons', obeying `any statistics' \cite{Wilczek}. The
restriction of two dimensions is important because then the flux tube cannot
be avoided in the exchange process, by crossing over in the third dimension.

Does this has any experimental basis? There are effective theories of
condensed matter phenomena where the concept of anyon has been found useful.
But, it does not have the same significance as the fundamental classification
into Bosons and Fermions. There are no fundamental anyons. Further, the flux
tube is not a physical entity. A charged particle that is executing an orbit
in the presence of a magnetic field has its dynamical phase $\int(p-eA)\cdot
dx/\hbar$. The magnetic flux contributes the extra phase $-e\nabla\times A(\pi
r^{2})/\hbar=-e\phi/\hbar$ in a ground state orbit with radius $r$. This has
to be a multiple of $2\pi$. Then $\phi=nh/e $. This is the flux quantization,
which is not a statement on the flux itself, but a requirement on the product
of the magnetic field and the area of the orbit. There is no meaning to the
flux tube without the real particle and the implicit particle orbit. There are
no fundamental flux quanta in physics independent of charged particles. The
usual magnetic field is not a collection of quantized flux tubes.

Now we present a mystery that will prove crucial in our discussion of the SSC
\cite{Unni-QHE}. Consider the cyclotron motion with frequency $\Omega=eB/m$ of
an electron in a 2D system, as a MOSFET used for observing the quantum Hall
effect. The wavefunction of the ground state Landau level in the applied
magnetic field has to obey the flux quantization condition $\phi=nh/e$. The
quantity $h/e$ is indeed the flux quantum in the quantum Hall effect. However,
the wavefunction also acquires a phase from the interaction of the magnetic
moment with the magnetic field, $\int\mu\cdot Bdt/\hbar=2\pi\mu_{B}%
B/\hbar\Omega=\pi$. Then the requirement of $2\pi$ closure phase on the orbit
is violated! Something is missing.

\section{Spin and Gravity - the missing link}

I assert that the reason why we have not found a satisfactory understanding
and proof of the spin-statistics connection is because we have not yet
considered the true physical identity of spin! \ We already learned a lesson
about mass and inertia, but it took 300 years to come to a proper
understanding of the equivalence of these two entities in physics. Mass is the
inertia for dynamics. Mass is also the charge of gravity; in fact that is its
paramount role.

\textit{Spin and angular momentum are, in all cases, the current of
matter-energy, or the current of the charge of gravity}. The role of spin and
angular momentum in gravitation is akin to the role of a magnetic moment in
electrodynamics. All physical effects involving a magnetic moment can be
traced to some electrodynamic effect. Since spin is the microscopic limit of
the closed current of mass-energy, the charge of gravity, all spin-dependent
physical effects should be traced ultimately to gravity, just as all physical
effects involving mass can be traced to gravity. The interaction of a magnetic
moment $\left(  \mu=\frac{1}{2}\int\left(  r\times j\right)  dV\right)  $ with
the magnetic field is $E=\mu\cdot B.$ The angular momentum or spin is $l=\int
r\times\bar{p}dV$ where $\bar{p}$ is the momentum density in terms of the
mass-energy density. \ Therefore, the coupling of the angular momentum is to a
`gravitomagnetic field' generated by the angular momentum or rotation of
matter. This field is the analogue of the magnetic field in relativistic
gravitation, with mass-energy currents as its source. The Lense-Thirring
precession effect (called `frame-dragging') in general relativity arises from
this. The coupling of the spin with the gravitomagnetic field is then
$E=l\cdot B_{g}/2.$ \ 

Most of our fundamental theories were completed well before any significant
knowledge about our cosmos was acquired, especially from an observational
point of view. Hence, most of our theories assume the `empty flat space-time'
background. In particular the theories of relativity explicitly assume an
empty space for their construction, which is blatant conflict with the factual
situation that was realized decades later. But, the theories have remained the
same, carrying this inconsistency. It is very easy to show and convince
oneself that the space and the `metric' of the matter-filled universe become
anisotropic in a frame moving relative to the average rest frame of all
matter-energy (identical to the preferred frame indicated by the cosmic
microwave background) \cite{Cosrel-ISI}. If that is the case, a theory based
on the invariance of the fundamental metric in all inertial frame is obviously
inconsistent. In fact, there is direct experimental evidence that the one-way
speed of light is Galilean ($c\pm v$), instead of being an invariant (the
Michelson-Morley experiment deals with two-way speed and second order effects)
\cite{Unni-nagpur}.%

\begin{figure}
[ptb]
\begin{center}
\includegraphics[
height=1.34in,
width=5.3889in
]%
{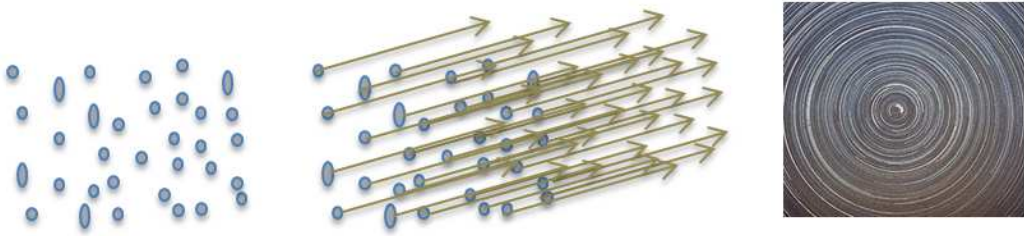}%
\caption{Space filled with matter cannot maintain its isotropy in a moving
frame. The anisotropy is linked to the matter-current that modifies the
gravitational potentials. Circular motion results in nonzero `curl' and a
`gravitomagnetic field'.}%
\end{center}
\end{figure}

One of the simplest calculations that one may do regarding the gravity of the
universe, knowing that it has a finite age of about 14 billion years, is an
estimate of the gravitational potential due to all the matter in causal
contact with us today. Though such a Newtonian concept is not rigorous in the
context of general relativity, it is instructive. Taking the average density
consistent with observations as $10^{-29}$ g/cm$^{3}$, we have
\begin{equation}
\Phi_{N}=\int_{0}^{R\approx cT}\left(  4\pi r^{2}drG\rho\right)  /r=2\pi G\rho
c^{2}T^{2}\approx c^{2}%
\end{equation}
since the quantity $2\pi G\rho T^{2}$ is approximately unity. Note that for
matter-dominated an radiation dominated evolutions, $\rho$ evolves as
$1/T^{2},$ and therefore $\Phi_{N}$ remains a constant at $c^{2}.$ This is
already remarkable since it is possible to make the theoretical claim that the
correct relativistic transformation for space and time involves the cosmic
gravitational potential and not the square of the speed of light as usually
assumed. Such a claim and the associated phenomenology are valid in the
context of all known experimental tests. A particle that is moving in the
cosmic gravitational potential $\phi$ of the universe experiences a modified
gravitational potential and a vector gravitational potential equal to
\begin{align}
\phi^{\prime}  &  =\phi(1-V^{2}/c^{2})^{-1/2}\\
A_{i}  &  =\phi\frac{V_{i}}{c}%
\end{align}
For universe with critical density, the quantity $\phi=1$ (in unit of $c^{2}
$). Circular motion then gives either a time dependent vector potential in
which the direction of potential changes, or a nonzero curl for the velocity
field, depending on the details of the motion. In the first case one gets an
electric-like gravitational effect, with no direct coupling to spin and in the
second case of pure rotations there is a gravitomagnetic field due to the
entire universe,%

\begin{equation}
\overrightarrow{B}_{g}=\nabla\times\mathbf{A}=2\omega
\end{equation}
This gravitomagnetic field couples to the spin angular momentum, analogous to
the coupling of the magnetic moment to a magnetic field in electromagnetism.
The coupling in the case of gravity is
\begin{equation}
H_{i}=\frac{1}{2}\left(  \vec{s}\cdot\overrightarrow{B}_{g}\right)
\end{equation}

Now, we will make this argument more rigorous. In a nearly isotropic and
homogenous universe with critical density, symmetry dictates the physical
metric as the Robertson-Walker metric,
\begin{equation}
ds^{2}=-c^{2}dt^{2}+a^{2}(t)(dr^{2}+r^{2}d\Sigma^{2})=-c^{2}dt^{2}%
+a^{2}(dx^{2}+dy^{2}+dz^{2})
\end{equation}
The time dependent scale factor $a(t)$ changes very slowly and can be
considered nearly a constant over time scales relevant to laboratory
experiments (or even the age of a typical successful theory).

\qquad This indicates a preferred frame with absolute time! Since there is
matter, there is a matter-current and anisotropy in a moving frame. This
anisotropic metric is obtained by transforming the Robertson-Walker metric
into the frame moving through the universe at velocity $v(t)$ (fig. 1). To
highlight the essential feature, we choose the direction of motion to be the
$x$ axis. Then, $x^{\prime}=x-vt,~t^{\prime}=t.$ Also, we redefine the
coordinates by absorbing the slowly varying scale factor ($\dot{a}%
/a\simeq10^{-18}m/s/m$) into the spatial coordinate labels. Then a
\emph{Galilean boost} is
\begin{align*}
x^{\prime}  &  =x-\left(  \frac{v}{c}\right)  ct\\
t^{\prime}  &  =t
\end{align*}

This is a coordinate transformation $(x,ct)$ of the privileged cosmic frame to
$(x^{1},x^{0}=x^{\prime},ct^{\prime})$ of a moving frame, representing
physical motion relative to cosmic matter. (It should be evident that a
Lorentz transformation that leaves the metric invariant is not physically
proper in nonempty space \cite{Cosrel-ISI}).The metric changes from that of
the FRW universe as a result of this motion, from the approximate $\delta
_{ik}$ to a new $g_{ik}.$ Therefore, the physical measures of time and space
in the moving frame are different from just the coordinate measures. We can
calculate the metric from the coordinate relations $x^{i}(x,t)$ as%

\begin{equation}
g_{ik}^{\prime}=%
\begin{bmatrix}
-\left(  1-v^{2}/c^{2}\right)  & -v/c\\
-v/c & 1
\end{bmatrix}
\end{equation}
$\allowbreak$

Other components are $\delta_{ik}^{\prime}$. \ The new non-zero metric
components in the moving frame are $g_{00}^{\prime}=-(1-v^{2}/c^{2}%
),~g_{01}^{\prime}=-v/c,$ and $g_{ii}^{\prime}=1.$ The components
$g_{01}^{\prime}=g_{10}^{\prime}=g_{tx}^{\prime}=-v/c$ in the frame moving
through the matter filled universe are in fact gravitomagnetic potentials in
the post-Newtonian language. Here $v$ is identical to the (absolute) velocity
that causes the Doppler dipole anisotropy of the CMBR. Therefore, there is
indeed an effective `gravitational vector potential' $A_{i}=-v_{i}$ in the
moving frame. Its curl, as experienced in every rotating frame, is the cosmic
gravitomagnetic field, $B_{g}=-\nabla\times v=2\Omega$.

I have shown elsewhere that cosmic gravity is the determining factor of both
relativity and dynamics, as described in theory of Cosmic Relativity
\cite{Cosrel-ISI}. There is a master determining frame defined by cosmic
matter-energy content and its gravity. The metric I derived already shows
clearly that relativistic effects like time dilation is already contained in
Galilean transformation! Numerous experimental results in mechanics,
relativity, electrodynamics, and propagation of light etc. give strong support
to the correctness of Cosmic Relativity \cite{Unni-nagpur}.

With this background, it is easy to see that all spin-dependent effects can be
consistently and correctly attributed to the spin-gravity interaction. This is
an important change of view in fundamental physics, where spin assumes its
true gravitational role. Spin can interact gravitationally with other matter
currents and the most important effects will occur in rotating frames in the
universe. In such frames, there is enormous matter current with circulation,
equivalent to a large gravitomagnetic field, and the interaction is $\left(
s\cdot\operatorname{curl}A_{g}\right)  /2=-\vec{s}\cdot\vec{\Omega}$. Spin can
be affected by forces $-s\times\Omega$ or a quantum phase can change by the
interaction, $\Delta\varphi=-s\cdot\Omega/\hbar$. The coupling is the same for
classical and quantum angular momentum. This is the physical reason why the
method of transforming to rotating frames generates physical effects akin to
the presence of a pseudo-magnetic field, provided there is a magnetic moment
associated with the spin, as in nuclear magnetic resonance
\cite{Rabi-Ramsey-Schwinger}. Factually, the interaction is between the spin
and the induced cosmic gravitomagnetic field, $-\vec{s}\cdot\vec{\Omega}$.
This then can be rewritten in terms of the magnetic moment, the gyromagnetic
ratio $g_{m}$ and an equivalent magnetic field $g_{m}\mu_{B}s\cdot B_{eq}$
where $B_{eq}=\Omega/g_{m}$. It is the phase changes that are relevant for the
spin-statistics connection.

Suddenly, we feel close to a truly fundamental reason for the SSC and Pauli
exclusion. I will show that the interaction of the spins of identical
particles with the relativistic cosmic gravitomagnetic field gives the exact
phases required for their characteristic difference, requiring symmetric
states for Bosons and anti-symmetric states for Fermions!

\section{Proof of the Spin-Statistics Connection}

Cosmic Relativity suggest that it is the gravitational interaction of the
quantum particles with cosmic matter that is responsible for the
spin-statistics connection \cite{Unni-spin}. In other words, the Pauli
exclusion is a consequence of the relativistic gravitational interaction of
the spin with the mass-energy in the universe. The quantum physical behavior
of two particles, as opposed to just one, is governed by their relative phase.
Consider the scattering of two identical particles, sketched in figure 2.%

\begin{figure}
[ptb]
\begin{center}
\includegraphics[
height=2.4242in,
width=4.6683in
]%
{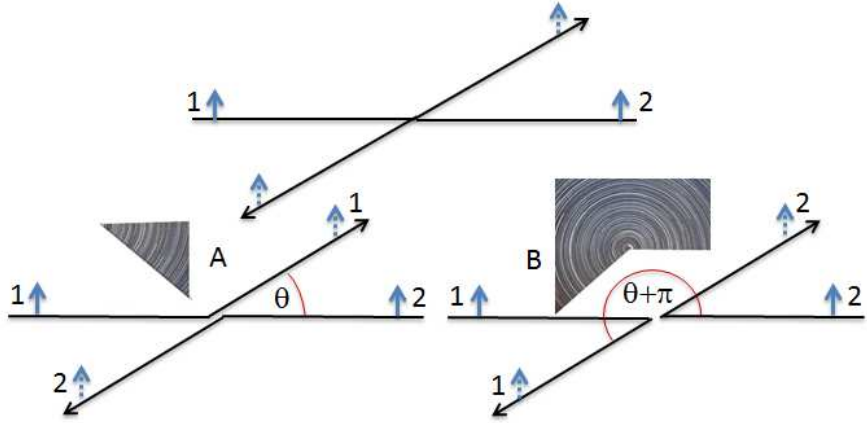}%
\caption{Interfering amplitudes for scattering between indistinguishable
particles with spin. The process in the upper panel involves two interfering
amplitudes, sketched in the lower panel. The difference between the two is the
angles through which the momentum vector rotates. In the presence of cosmic
matter, scattering induces the gravitomagnetic potentials for the duration of
such rotation, schematically indicated by the matter rotating in the centre of
mass frame. Therefore, there is a definite gravitational phase difference
$2s(\delta\theta)=2s\pi$ between the two amplitudes.}%
\end{center}
\end{figure}

The process can happen by two quantum amplitudes, shown in the lower panel.
The amplitudes for these two processes differ by only a phase for identical
particles in identical states, since the initial and final states in the two
processes are indistinguishable. The particles are assumed to be spin
polarized in identical directions, perpendicular the plane containing the
scattering event. Only then, the initial and final states are
indistinguishable, and interfering. We can calculate the phase changes in any
configuration, but at present we want to discuss only the phase difference for
indistinguishable states. The two processes are different in the angle through
which the momentum vectors of the particles turn. In fact that is the only
difference between the two amplitudes. The difference in angles is just $\pi$.
(This is why it is equivalent to an exchange - what is really exchanged is the
momentum vector after the scattering). The dynamics happens always in the
gravitational potential of the entire universe. The calculation for each
particle can be done by noting that the k-vector can be considered without
deflection, but the entire universe turned through an appropriate angle, with
angular velocity of this turning decided by the rate of turning of the
k-vector. It does not matter whether we are dealing with massless particles or
massive particles since the only physical fact used is the change in the
direction of the k-vector of the particle.

As discussed earlier, the motion of the particle relative to the universe
generates the vector potential, and the rotation of the momentum vector
generates a nonzero curl and therefore a gravitomagnetic field,
$\overrightarrow{B}_{g}=\nabla\times\mathbf{v}(t)=2\overrightarrow{\Omega}(t)$
\ where $\overrightarrow{\Omega}$ is the rate of rotation of the $k$-vector.
This field exists only for the duration of the turning of the $k$-vector. The
change in phase is
\begin{equation}
\Delta\varphi_{g}=\frac{1}{\hbar}\int(\vec{s}\cdot\vec{\Omega})dt=\frac
{1}{\hbar}\int(\vec{s}\cdot\frac{d\vec{\theta}}{dt})dt=\frac{\vec{s}\cdot
\vec{\theta}}{\hbar}%
\end{equation}
The phase depends only on the angle through which dynamical path turns. We see
that, remarkably, this additional phase is independent of the duration $t $.
This is the reason that there is an apparent connection to geometric phases,
but in reality the quantum dynamical phase arises in the gravitational interaction.

As an important interlude, we see that the phase due to the interaction of the
spin with the motion-induced cosmic gravitomagnetic field in a cyclotron
Landau orbit that we discussed earlier is $\varphi_{g}=2\pi s=\pi$. This, when
added (with either sign) to the equal phase from the magnetic moment-magnetic
field interaction, we get $2\pi$ (or $0$); only then the correct closure phase
for the orbit is obtained \cite{Unni-QHE}. \emph{This proves the reality of
cosmic gravitational potential and the verity of the spin- gravitomagnetic
field interaction}.

The momentum vectors turn in the same sense for both the particles (the
product wavefunction) and therefore the total phase change is $\varphi
_{1}=2s\theta_{1}/\hbar$ where $\theta_{1}$ is the angle through which the
$k$-vector turns for the first amplitude. For the second amplitude the phase
change is $\varphi_{2}=2s\theta_{2}/\hbar$. The \textit{phase difference
between the two amplitudes} is
\begin{equation}
\Delta\varphi=2s(\theta_{1}-\theta_{2})/\hbar=2s/\hbar\times\pi
\end{equation}

The rest of the proof of the spin-statistics connection is straightforward.
For zero-spin particles the proof is trivial since
\begin{equation}
\Delta\varphi=s(\theta_{1}-\theta_{2})=0\times2\pi=0
\end{equation}
and therefore zero-spin particles are Bosons and their scattering amplitudes
add with a $+$ sign. \ Zero-spin particles have no spin-coupling to the
gravitomagnetic field of the universe and there is no phase difference between
the two possible amplitudes in scattering. What remains is the proof for
spin-half particles, since the higher spin cases can be constructed from
spin-half using the Schwinger construction (both the phase and the spin
projection for identical states is just additive). The phase difference
between the two amplitudes then is%
\begin{equation}
\Delta\varphi=2s(\theta_{1}-\theta_{2})=2\times\frac{1}{2}\times\pi=\pi
\end{equation}
The relative sign between the amplitudes is $\exp(i\pi)=-1.$ The amplitudes
add with a negative sign. \textit{Therefore, the half-integer spin particles
obey the Pauli exclusion and the Fermi-Dirac statistics.} The exchange of
\emph{any} two particles in a multiparticle system of identical fermions
introduces a minus ($-$) sign between the original amplitude and the exchanged amplitude.

Particles of higher spins can be considered (for this proof) as composites of
spin-1/2 particles. When the spin projection is integer valued $\left(
n\hbar\right)  $, we have $\Delta\varphi=2s(\theta_{1}-\theta_{2})=2n\pi.$
Therefore, integer spin particles obey Bose-Einstein statistics.

The proof is valid for interacting particles since the phase changes due to
interactions are identical for the two amplitudes, as all other dynamical
phases in the relevant diagrams.

This remarkable connection between quantum physics and gravity of the universe
is indeed startling. But this cosmic connection is also a natural consequence
in a critical universe in which everything is gravitationally interacting with
everything else. It is satisfying to see that a deep physical phenomenon is
linked to a universal physical interaction and not just to mathematical
structures and consistency. This reaffirms my conviction that every physical
change, including changes in quantum phase, must occur through a physical
interaction involving one of the four interactions (we know of at present). We
can now calculate the gravitational phase change for the two amplitudes for
any \emph{arbitrary initial and final states}, and thus we have the most
general statement about the relative phase between multi-particle quantum amplitudes.

\section{Epilogue}

The first `oriental' and English translation of Einstein's general theory of
relativity was by S. N. Bose and M. N. Saha, in 1920, as very young lecturers
in physics \cite{Saha-Bose}. Bose's last published works pertain to Einstein's
attempt at a unified field theory, based on the mathematical deviation from
general relativity with a non-symmetric metric tensor. But, generally
speaking, Bose didn't show much interest in the conceptual and fundamental
issues in gravity or quantum mechanics, except in that one decisive occasion
of his derivation of the Planck law of radiation, and an immediate follow up
paper, both translated for publication by Einstein. He was perhaps drawn more
by the mathematical aspects in theoretical physics \cite{Blanpied}. S. N. Bose
would have been very surprised to see the role of gravity in the statistical
mechanics of light quanta he derived at a time when nothing significant about
the universe and its matter-energy content was known. Observational cosmology
matured only after the protagonists involved in the development of the quantum
mechanics of spin all left the stage. By then, relativity theories based on
the unreal and non-existent empty space had taken deep roots, giving the
impression of a robust framework. Despite that well known criticism of the
physics of inertia and dynamics based in empty-space by Ernst Mach, the rapid
cruise of physics continued with any course correction. The obvious reality of
the gigantic cosmic gravitational potentials was not noticed for a whole
century, while observational cosmology slowly matured. Today, we are in a
position to see the truth of the situation from many experimental evidence,
old and new. Cosmic gravity is the decisive factor for relativity and
dynamics. A fundamental problem like the origin of the spin-statistics
connection involving the current of mass-energy as the `spin' of fundamental
particles could not have been solved without invoking gravity. Consideration
of phase changes of quanta with spin in the ever present cosmic gravitational
fields yielded the surprising and pleasing result that the fundamental
connection between spin and statistics in quantum theory is a consequence of
the gravitational interaction of the spin with the matter in the universe.
This is a deep result, valid in very general physical situations. It is simple
and transparent to understand. Atoms and matter of wide variety exist stably
because Fermions obey Pauli exclusion. Wonders like lasers are possible
because Bosons tend to bunch together. At a deeper layer, it is cosmic gravity
that provides the basis for this rich variety of phenomena. As in collapsing
stars, only gravity can violate what has been set up by gravity.

\section{Acknowledgments}

In fond memory of E. C. George Sudarshan, teacher and friend, who understood
and helped us understand Bosons and Fermions through his phenomenal work in
quantum optics and the theory of weak interactions.

\end{document}